\documentclass[a4paper, 12pt]{article}
\usepackage{epsfig}
\usepackage{graphicx}
\usepackage{amsmath}

\begin{document}

\title{Weierstrass's criterion and \\
compact solitary waves}

\author{Michel Destrade, Giuseppe Gaeta, Giuseppe Saccomandi}

\date{2007}

\maketitle

\begin{abstract} Weierstrass's theory is a standard qualitative tool for single degree of
freedom equations, used in classical mechanics and in many
textbooks. In this note we show how a simple generalization of
this tool makes it possible to identify some differential
equations for which compact and even \emph{semi-compact}
travelling solitary waves exist. In the framework of continuum
mechanics, these differential equations correspond to bulk shear
waves for a special class of constitutive laws.
\end{abstract}
 \section{Introduction}

A \emph{compact wave} is a robust solitary wave with a compact
support, outside of which it vanishes identically. A
\emph{compacton} is to a compact wave what a soliton is to a
solitary wave with an infinite support; that is, a compacton is a
compact wave that preserves its shape after colliding with another
compact wave. Rosenau and Hyman (1993) introduced these concepts
over a decade ago and a substantial number of differential
equations supporting compact waves has since been identified and
studied.

From the mathematical point of view, the emergence of such
solutions is related to the degeneration of the differential
equations of motion at certain points, and to the corresponding
failure of the uniqueness theorem at these. Indeed, compact waves
are a patchwork made pasting together at degenerate points the
different possible solutions (unique in between degenerate
points), hence patching together non-unique solutions; it follows
that they are not analytic solutions, and in this respect they are
substantially different from standard soliton solutions.

Compact waves are weak solutions of differential equations, which
are continuous -- in contrast to shock waves -- but have
discontinuous derivatives, similarly to acceleration waves. A
simple introduction to compact waves and to compactons can be
found in a recent article by Rosenau (2005).

The main objection facing compact waves is that the link between
an \emph{adequate} partial differential equation and a
\emph{constitutive} law is often tenuous. Indeed, the generic
affirmation that compact waves emerge from a balance between
higher nonlinearity and nonlinear dispersion remains vague and
esoteric if it is not supported by a clear and rigorous mechanical
derivation of the \emph{right} equations.

Destrade and Saccomandi (2006$^a$, 2006$^b$) recently proposed a
general theory of dispersive nonlinear acoustics and showed how it
is possible to derive exact equations governing the propagation of
compact shear waves in solids with an inherent characteristic
length. By applying standard asymptotic procedures to these exact
equations, it is then possible to justify evolutions equations
which are similar to the compacton factory known as the $K(m,n)$
\emph{KdV equation}.

From the mechanical point of view, these
results were obtained by a careful modelling of the dispersive
part of the Cauchy stress tensor.
The corresponding constitutive equations unify and explain in
depth several theories of weakly \emph{nonlocal} continuum
mechanics, such as the $\alpha$-LANS theory of turbulence, or the
Rubin, Rosenau, and Gottlieb (2005) theory of inherent
characteristic length.

From the mathematical point of view, the emergence of compact
waves is investigated by a simple modification of Weierstrass's
theory (a useful qualitative tool for single degree of freedom
equations in classical mechanics).

In this note we show how another class of constitutive assumptions
can generate the emergence of compact waves and even of
\emph{semi-compact waves}, which are travelling solitary waves
with a semi-infinite support.

\section{A generalization of Weierstrass's discussion}

One of the most elegant and powerful tools for the qualitative
analysis of one-dimensional Lagrangian conservative motions is
\emph{Weierstrass's theory}. When an integral of energy (in a
generalized sense) exists, Weierstrass's theory allows us to
understand whether the motion of our Lagrangian system is periodic
or non-periodic, simply by looking at the roots of the potential
function.

Let us imagine that our natural Lagrangian system can be described
by a single holonomic parameter $q$, say. Furthermore, suppose
that it is time-independent. The energy integral of such a system
is
\begin{equation}
T(q,\dot{q}) - U(q) \ = \ E \ ,  \label{p0}
\end{equation}
where $T$ is the kinetic energy, $U$ is minus the potential energy
(note the unconventional choice of sign), and the constant $E$ is
determined by initial conditions.

Recall that for natural Lagrangian systems, the most general form
of the kinetic energy is $T = a(q) \dot{q}^2/2$, where $a=a(q)$ is
a positive function of $q$ only; also, $T = 0$ only if
$\dot{q}=0$. We will thus assume (\ref{p0}) is in the form
\begin{equation}
\frac{1}{2} \ a(q) \, \dot{q}^2 \ - \ U(q) \ = \ E \ . \label{p1}
\end{equation}

It follows that our motion is confined to those configurations
where $U(q) + E \geq 0$. The special configurations $\overline{q}$
(say) where $U(\overline{q}) + E=0$ and $U' (\overline{q}) \not=
0$ are called \emph{barriers} because they cannot be crossed by
the motion of $q(t)$: they split the range of possible values for
$q$ into allowed and prohibited intervals. Points with
$U(\overline{q}) + E=0$ and  $U' (\overline{q}) = 0$ are
\emph{soft barriers} and separate two allowed intervals (see
below).

Solving first \eqref{p1} for $\dot{q}$, and then differentiating with respect to $t$,
we find in turn
\begin{equation}
\dot{q}^{2} \ = \ 2 \, \dfrac{U(q) + E}{a(q)}, \quad
 \ddot{q} \ = \ \dfrac{1}{a(q)}U'(q) \, - \,
\dfrac{a'(q)}{a^2(q)} \, [U(q) + E] \ . \label{p3}
\end{equation}
Hence at a barrier $\bar{q}$ which is a \emph{simple root} of
$U + E = 0$ (that is, such that $U(\overline{q}) + E = 0$ and $U'(\overline{q}) \ne 0$),
we have $\ddot{q} \ne 0$.
Such a barrier is called an \emph{inversion point},
because the motion reverses its course after reaching it.

We now separate the variables in \eqref{p3}$_1$ and integrate to find
\begin{equation}
t \ = \ \pm \int{\sqrt{\dfrac{a(q)}{2[U(q) + E]}} \text{d} q} \ .
\label{integrate}
\end{equation}
At a (soft) barrier which is a double root of $U+E = 0$ (that is,
such that $U(\overline{q}) + E = U'(\overline{q}) = 0$), the
integral diverges. Thus a soft barrier is also called an
\emph{asymptotic point} because it takes an infinite time to reach
it.

\section{Wave propagation}

Now we turn to wave propagation.
Consider the case of \emph{semi-linear wave equations} in the
unknown field $u=u(x,t)$,
\begin{equation}
u_{tt} \ - \ c^2 \, u_{xx} \ = \ F(u) \ ,  \label{p4}
\end{equation}
where $c$ is a constant and $F$ a nonlinear function of $u$. Then
for travelling waves of speed $v$, i.e. for $u$ in the form
\begin{equation}
u(x,t) \ = \ \varphi(z) \ , \ \qquad z := x - v t,  \label{p5}
\end{equation}
we obtain the second order differential equation
\begin{equation}
(v^2 - c^2) \varphi'' - F(\varphi) = 0.  \label{p6}
\end{equation}
Multiplying by $\varphi'$ and integrating, we find an equation of
the same form as the first equation in \eqref{p3}, where $a$ is
the constant $v^2 - c^2$ and $U$ is the anti-derivative of $F$;
note that $E$ is now related to the integration constant.

Weierstrass's theory tells us that a periodic wave corresponds to
the existence of two consecutive inversion points; that a pulse
solitary wave with infinite tails corresponds to the existense of
an asymptotic point followed by an inversion point; and that a
kink solitary wave with infinite tails corresponds to two
consecutive asymptotic points; see Peyrard and Dauxois (2004) or
Kichenassamy and Olver (1992) for similar discussions.

Consider the case of the following fully \emph{non-linear wave equations},
\begin{equation}
u_{tt} \, - \, c^2 \, u_{xx} \, - \, c_\text{NL}^2 \, (u_x^3)_x \
= \ F(u) \ , \label{p7}
\end{equation}
where $c_\text{NL}$ is a constant. The travelling wave reduction
\eqref{p5} yields
\begin{equation}
(v^2 - c^2)\, \varphi'' \ - \ c_\text{NL}^2 \, [(\varphi ')^3]' \
= \ F (\varphi) \ . \label{p8}
\end{equation}
Here Weierstrass' discussion must be modified, sometimes to
dramatic effect.

Take for instance the degenerate case $v^2 = c^2$. In that case, a
first integral of \eqref{p8} is
\begin{equation}
(\varphi')^4 \ = \ 2 \ \frac{U(\varphi) + E}{a} \ , \label{p9}
\end{equation}
where $U(\varphi) \equiv -\int{F}$, $E$ is a constant of
integration, and $a = 3 c_\text{NL}^2 /2$.

We can still conduct an analysis \emph{\`a la} Weierstrass, but we
find that the fourth-order power above introduces some new
features, not present in mechanical conservative Lagrangian
systems. Indeed, now the barriers $\overline{\varphi}$ are
attainable not only when they are simple roots of $U + E = 0$, but
also when they are double roots. This is the case because the
analogue to \eqref{integrate} is here
\begin{equation}
z \ = \ \pm \, \int{\sqrt[4]{\dfrac{a}{2[U(\varphi) + E]}} \
\text{d} \varphi} \ , \label{integrate2}
\end{equation}
and the integral converges for double roots.

Saccomandi (2004) gives a detailed discussion on this possibility
and explains its consequences by the failure of the Lipschitz
condition, leading to a possible lack of uniqueness, and
eventually to the possibility of compact waves; see Destrade and
Saccomandi (2006$^a$, 2006$^b$) for examples.

In the present note we study yet another possibility for
Weierstrass's theory, touched upon by Ferrari and Moscatelli
(1997), Rosenau (2000), and Gaeta, Gramchev and Walcher (2006);
namely the case of barriers which are roots of \emph{non-integer
order} to the equation $U + E = 0$.

For a first glimpse at what may happen in this case, we consider
in turn two examples with roots of fractionary order.

\bigskip\noindent
{\tt Example 1.} In the first example, we take $E = 0$, $a=2$, and
\begin{equation} U(q) \ = \ \sqrt{q} \ (1 - \sqrt{q}) \ . \end{equation}
Here we record two barriers: $\overline{q}_1 = 0$ and
$\overline{q}_2 = 1$. The integral in \eqref{integrate} yields
\begin{equation}
t \ = \ - \, 2 \ \sqrt{\sqrt{q} - q} \ + \ \arcsin(2\sqrt{q} - 1)
\ . \label{p12}
\end{equation}
Clearly, both barriers can be reached in finite times: with our
choice of the integration constant, these are $t= - \pi/2$ for
$\overline{q}_1$, and $t=0$ for $\overline{q}_2$. From \eqref{p3}
the acceleration is
\begin{equation}
\ddot{q} = \dfrac{1 - 2\sqrt{q}}{2\sqrt{q}}.  \label{p13}
\end{equation}
At $\overline{q}_2 = 1$ the acceleration is not zero, and the
motion reverses; at $\overline{q}_1 = 0$ however, the acceleration
blows up!

\begin{figure}
\centering
\includegraphics[width=80mm, height=50mm]{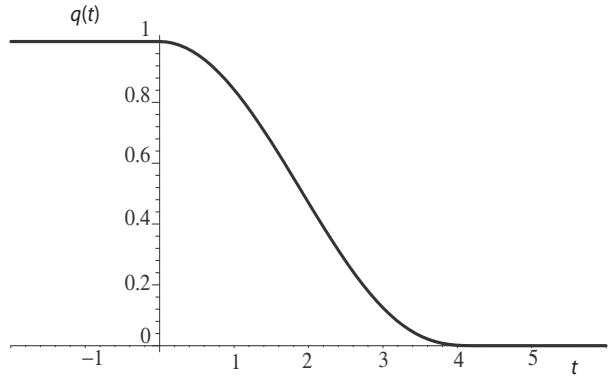}
\caption{Example 2. Compact solitary kink wave, patched together
using fractionary roots in Weierstrass's discussion, with a
potential of the form: $U(q)=q^{\frac{4}{3}}(1-q^{\frac{1}{3}})$.
\label{figure_kink}}
\end{figure}

\bigskip\noindent
{\tt Example 2.} In the second example, we take $E = 0$, $a=2$,
and
\begin{equation} U(q) \ = \ q^{4/3} \ (1 \, - \, q^{1/3}) \ . \end{equation}
Here we record two barriers as well, again $\overline{q}_1 = 0$
and $\overline{q}_2 = 1$. We perform the integral in
\eqref{integrate} for $t \in [-3\sqrt{2}, 3\sqrt{2}]$ and solve it
explicitly for $q$ as
\begin{equation}
q(t) \ = \ \left(1 - \frac{t^2}{18} \right)^3 \ .  \label{p15}
\end{equation}
Clearly again, the barriers are reached in finite times:
$\overline{q}_1 = 0$ at $t=\pm 3\sqrt{2}$ and $\overline{q}_2 = 1$
at $t = 0$. We also find that the acceleration is given by
\begin{equation}
\ddot{q} = \dfrac{1}{3} q^{\frac{1}{3}}(4 - 5q^{\frac{1}{3}}),  \label{p16}
\end{equation}
making it clear that the barrier $\overline{q}_2 = 1$ is a
configuration associated with a finite force, whereas the barrier
$\overline{q}_1 = 0$ is associated with an equilibrium.
Furthermore, note that the right hand-side of \eqref{p16} does not
satisfy the Lipschitz condition at $q=\overline{q}_1 = 0$. This
allows non-uniqueness of solution, and in fact we can patch
together a compact solitary kink: this is equal to one for $t <
0$, then for $0 < t < 3\sqrt{2}$ it is accelerated to negative
velocity and decreases to reach zero at time $t = 3\sqrt{2}$, and
then stays there for $t>3\sqrt{2}$; see Figure \ref{figure_kink}.

\bigskip
These two examples highlight the complexity and the richness of
the situations arising when considering non-integer barriers.

Let us now turn to a more general (non-fractionary) case. We take
$E = 0$, $a=2$, and
\begin{equation}
U(q) \ = \ q^{2r} \ V(q^2) \ ,  \label{p17}
\end{equation}
where $V$, and thus $U$, have a barrier $\overline{q} > 0$
($V(\overline{q}^2) = 0$). Then by \eqref{p3}$_2$, the
acceleration is
\begin{equation}
\ddot{q} = r q^{2r-1} V(q^2) + q^{2r + 1} V'(q^2).
\label{p18}
\end{equation}
We should consider several cases, depending on the value of $r>0$.

If $2r < 1$, then $\ddot{q}$ blows up at the barrier $q
\rightarrow 0$, similarly to the first example above. For this
case we conclude that sublinear ($2r < 1$) roots  correspond to
singular points which cannot be reached in any way.

If $2r = 1$ or $2r = 2$, we recover the already discussed cases of
simple and double roots.

If $2r \ge 2$, the integral in \eqref{integrate} diverges and the
barrier at zero cannot be reached in a finite time -- it is an
asymptotic point.

If $1 < 2r < 2$ however, the integral is finite and the barrier
zero is reached in a \emph{finite} time. Also, by \eqref{p18},
$\ddot{q} = 0$ at that barrier, and we thus \emph{arrive at an
equilibrium configuration with null velocity}, a situation which
gives rise to a predicament: will the motion settle on this
equilibrium configuration \emph{ad infinitum} or will it reverse
its course?

In fact, the Cauchy problem is ill-posed here, and the
accompanying lack of uniqueness gives us some latitude to patch
together compact waves (similarly to the second example above) or
semi-compact waves, as is seen in the next section.

\section{Semi-compact shear strain waves}

Destrade and Saccomandi (2006$^a$, 2006$^b$) show that the motion
of transverse strain waves in nonlinear dispersive solids is
governed by the following equation,
\begin{equation} \label{shear_wave}
\left( \mu W \right)_{xx} + \left( \alpha W_{tt} \right)_{xx} = \rho W_{tt},
\end{equation}
where $x$ is the direction of propagation, $W$ is the transverse
strain, and $\rho$ is the mass density. The constitutive
parameters are $\mu = \mu(W^2)$, the generalized \emph{shear
modulus of nonlinear elasticity}, and $\alpha$, the
\emph{dispersion parameter} of Rosenau et al. (1995). For
simplicity here, $\alpha$ is taken constant and the strain wave is
linearly polarized, travelling with speed $v$, see \eqref{p5}.
Then $W = W(x- v t)$ and \eqref{shear_wave} leads to
\begin{equation} \label{shear_wave_2}
\mu W + \alpha  v^2 W'' = \rho v^2 W.
\end{equation}

Next we write $\mu$ in the form
\begin{equation} \label{shear_modulus}
\mu(W^2) = \mu_0 - (\alpha \mu_0/\rho)
 \left[ r W^{2r - 1}V(W^2) + W^{2r + 1}V'(W^2) \right],
 \end{equation}
where $\mu_0$ is the ground state shear modulus, $r > 1/2$, and $V$
is an as yet arbitrary function.
For bulk waves, $v$ is arbitrary and here we fix it at the sonic speed
$v \equiv \sqrt{\mu_0 / \rho}$.
Then \eqref{shear_wave_2} is exactly of the same form as \eqref{p18}.

We are thus entitled to consider the possibility of barriers of non-integer order for the wave.
For instance, take $V$ in the form
\begin{equation} \label{V}
V (W^2 ) =  \gamma \frac{\rho}{\alpha}  \left( J_m - W^2 \right)^n,
\end{equation}
where $\gamma > 0$, $J_m >0$, and $n>1$ are constants.
Integrating \eqref{shear_modulus} with this choice gives
\begin{equation}
W^{'2} = \gamma \frac{\rho}{\alpha}
  W^{2 r}\left( J_m - W^2 \right)^n.
\end{equation}
Then the following change of variable and rescaling of function
\begin{equation}
\xi \equiv \left[  \gamma \frac{\rho}{\alpha} J_m^{r+n-1}
\right]^\frac{1}{2} z, \qquad \omega (\xi) \equiv
W([z(\xi)]/\sqrt{J_m},
\end{equation}
give the following non-dimensional version of the governing equation,
\begin{equation} \label{omega}
\dot{\omega}^2 = \omega^{2r}\left(1-\omega^2\right)^n.
\end{equation}

Now we briefly discuss whether \eqref{shear_modulus}-\eqref{V} constitutes a
reasonable shear response for a nonlinear solid.
We recall that the shear stress $\tau$ (say) necessary to maintain
a solid in a static state of finite shear with amount of shear $K$ (say)
is $\tau = \mu(K^2) K$, given here by
\begin{equation} \label{shear_stress}
\dfrac{\tau}{\mu_0} = K - \gamma \left[ r K^{2 r} \left(J_m - K^2 \right)^n
  - n K^{2 r + 2} \left(J_m - K^2 \right)^{n-1} \right].
\end{equation}

First we see that $K$ must not be allowed to go too far beyond
$\sqrt{J_m}$ for $\tau$ to remain positive. In practice, this
means that for a given material we must fix $J_m$ beyond the
maximal shear allowed before its rupture. Note that the actual
value of $J_m$ has no bearing on the existence and characteristics
of the shear wave because it does not appear in \eqref{omega}.

Second we remark that the graph of $\tau(K)$ and the graph of
$\mu_0 K$ (corresponding to a material with a linear shear
response) cross -- independent of $\gamma$ -- for $K = 0$, $K =
\sqrt{r J_m / (r+n)}$, and $K = \sqrt{J_m}$. The slopes of the
$\tau(K)$ graph at $K = 0$ and at $K = \sqrt{J_m}$ are defined
(and equal to $\mu_0$) when
\begin{equation} \label{restr}
2r > 1, \qquad \text{and} \qquad n>2.
\end{equation}
(Of course these slopes are also defined at $2r = 1$ and $n = 2$,
but we leave those special cases aside because they lead to simple
and double roots in Weierstrass's discussion, already treated
above.) The slope of the $\tau(K)$ plot at $K = \sqrt{r J_m /
(r+n)}$ is always greater than $\mu_0$. It follows that between
$K=0$ and $K= \sqrt{r J_m / (r+n)}$, the solid is strain-softening
in shear, and that between $K= \sqrt{r J_m / (r+n)}$ and $K=
\sqrt{J_m}$, the solid is strain-hardening in shear.

Third we make sure that the shear response is a monotone
increasing function of $K$, by taking $\gamma$ small enough so
that the equation $\tau'(K) = 0$ has no root in the $[0,
\sqrt{J_m}]$ interval. Note that the actual value of $\gamma$ does
not affect the non-dimensional equation of motion \eqref{omega}.

\begin{figure}
\centering
\includegraphics[width=85mm, height=50mm]{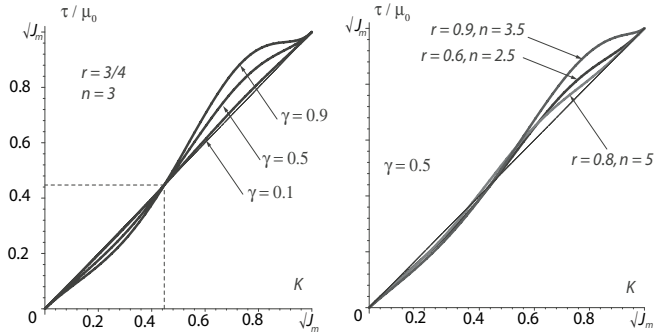} 
\caption{Shear stress response obtained by varying the
constitutive parameters in \eqref{shear_stress}. The plots cross
the linear stress shear response (thin straight line) at $K=0$,
$K= \sqrt{r J_m / (r+n)}$ (indicated by dashed lines on left
graph), and $K= \sqrt{J_m}$.\label{figure_shear}}
\end{figure}

Figure \ref{figure_shear} displays some examples of shear stress
responses satisfying the requirements just evoked.
 
Having checked that the shear response is sound, we may now look
for solutions to the non-dimensional equation \eqref{omega}
describing solitary wave solutions to the original equation
\eqref{shear_wave}; these are homoclinic or heteroclinic solution
to \eqref{omega}.

\begin{figure}
\centering
\includegraphics[width=80mm, height=50mm]{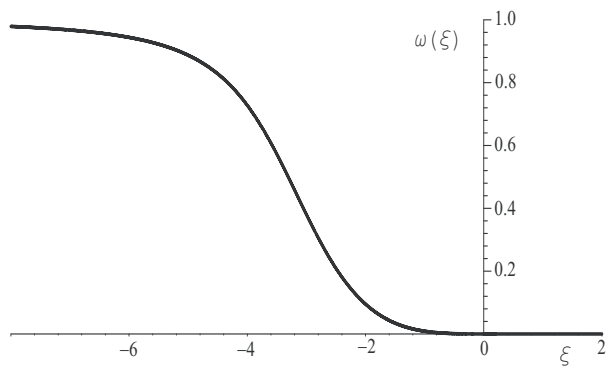}
\caption{Semi-compact solitary kink wave in a dispersive nonlinear solid
with shear stress response given by \eqref{shear_stress} at $r = 3/4$, $n = 3$.}
\label{figure_semi_kink}
\end{figure}

Owing to \eqref{restr}, the barrier $\overline{\omega} = 1$ is
necessarily an asymptotic point. Taking $r \ge 1$ also creates an
asymptotic point barrier $\overline{\omega} = 0$, leading to a
solitary kink with tails of infinite extend. However, taking $1/2
< r < 1$ gives a barrier reachable in a finite time. The result is
a \emph{semi-compact wave}, coming from value 1 at $- \infty$ and
decreasing to zero, which it reaches in a finite time with zero
speed and zero acceleration. It may then remain at this value
zero. For Figure \ref{figure_semi_kink} we took the case $r =
3/4$, $n = 3$, and chose $\omega(0) = 0$ to fix the value of the
integration constant.

\end{document}